\documentclass[pre,twocolumn,superscriptaddress,amsmath,amssymb,showpacs]{revtex4}

\usepackage{bm}

\usepackage[dvips]{graphicx}
\usepackage[dvips]{color}
\usepackage{hyperref}
\hypersetup{
colorlinks=true,
linkcolor=black,
citecolor=black,
filecolor=black,
urlcolor=black,
pagecolor=white
}
\newcommand{\grad}{\bm{\nabla}}
\newcommand{\vct}[1]{\mathbf{#1}}

\newcommand{\Vext}[1]{V_{\text{ext}}^{(#1)}}
\newcommand{\Vol}{\Gamma}
\newcommand{\Mob}{\gamma}

\begin{document}
\title{Selectivity in binary fluid mixtures: static and dynamical properties}

\author{Roland Roth}
\email{Roland.Roth@mf.mpg.de}
\affiliation{Max-Planck-Institut f{\"u}r Metallforschung,
Heisenbergstr. 3, 70569 Stuttgart, Germany} \affiliation{Institut
f{\"u}r Theoretische und Angewandte Physik, Universit{\"a}t Stuttgart,
Pfaffenwaldring 57, 70569 Stuttgart, Germany}
\author{Markus Rauscher}
\email{Rauscher@mf.mpg.de}
\affiliation{Max-Planck-Institut f{\"u}r Metallforschung,
Heisenbergstr. 3, 70569 Stuttgart, Germany} \affiliation{Institut
f{\"u}r Theoretische und Angewandte Physik, Universit{\"a}t Stuttgart,
Pfaffenwaldring 57, 70569 Stuttgart, Germany}
\author{Andrew J.\ Archer}
\email{A.J.Archer@lboro.ac.uk}
\affiliation{Department of Mathematical Sciences, Loughborough University,
Leicestershire, LE11 3TU, UK}

\date{\today}

\begin{abstract}
Selectivity of particles in a region of space can be achieved by applying
external potentials to influence the particles in that region. We investigate
static and dynamical properties of size selectivity in binary fluid mixtures
of two particles sizes. We find that by applying an external potential that is
attractive to both kinds of particles, due to crowding effects, this can lead
to one species of particles being expelled from that region, whilst the other
species is attracted into the region where the potential is applied. This
selectivity of one species of particle over the other in a localized region of
space depends on the density and composition of the fluid mixture. Applying an
external potential that repels both kinds of particles leads to selectivity of
the opposite species of particles to the selectivity with attractive
potentials. We use equilibrium and dynamical density functional theory to
describe and understand the static and dynamical properties of this striking
phenomenon. Selectivity by some ion-channels is believed to be due to this
effect.
\end{abstract}


\maketitle

\section{Introduction}

Biological ion channels are amazing nanofluidic devices. Ion channels are
special proteins with pores through the center that allow for passive
transport of ions, such as K$^+$ and Na$^+$, along their electrochemical
gradients, through a membrane formed by a lipid bilayer. The pores of these
proteins have a diameter of order a few \AA\/ngstroms and, in
general, have a charge of their own. Ion channels have two
important functions: (i) They can open and close the pore and thereby control
the current through the channel. This phenomenon is called gating. (ii)
Typically, they can select the type of ions that can pass through the pore, a
phenomenon called selectivity. Selectivity can occur with respect to several
properties. Some channels can select divalent ions over monovalent ones. Other
channels can distinguish ions of the same valency by size. Using gating and
selectivity, ion channels are responsible for a wide range of physiological
phenomena such as the regulation of ion concentrations inside the cell
\cite{hille01}. 

The selectivity with respect to ions of equal electrical charge (e.g., between
K$^+$ and Na$^+$) is rather puzzling. Recent theoretical studies of ion
channels with wide pores, such as the l-type Ca channel, examined the effects 
of entropy (also referred to as molecular crowding) in the three component
mixture composed of neutral solvent molecules and two ion species, which
differ only in size, and also studied the influence of the electrostatic
attraction of the ions into the channel \cite{roth05a}\/. It was found that
the entropic effects can dominate over the electrostatic attractions. The
theories used in these studies were (i) equilibrium density functional theory
(DFT) for mixtures of hard spheres in an external field and (ii) bulk fluid
models of charged hard spheres, in which the external potentials are mapped on
to shifts in the chemical potentials of the different components 
\cite{Nonner00,Nonner01,Gillespie02}.

In recent years, colloidal suspensions have become popular model systems
for testing all aspects of liquid state theory, including DFT. In contrast to
atomic or molecular fluids, colloidal suspensions can conveniently be studied
optically, e.g., by confocal microscopy. In addition, individual or groups of
colloids can be manipulated using optics: particles which are optically denser
than the surrounding solvent are attracted to regions of higher light
intensity, e.g, to a laser focus. This is the working principle of optical
tweezers, which allows for the creation of an attractive potential well for
the colloids, by means of laser irradiation. Because of this ability to
manipulate and observe the individual colloids, we believe that, as has been
the case for other liquid state phenomena, colloidal suspensions provide a
useful model system in which to understand how selectivity in some ion channels
occurs. Thus, although the present study is motivated by the desire to further
understand entropy driven selectivity in ion channels, we consider a more
general scenario. Using an accurate DFT, we study equilibrium selectivity
and using dynamical density functional theory (DDFT) we study the dynamics of
selectivity in colloidal binary mixtures. 

Dynamical aspects of selectivity have neither been studied experimentally nor
theoretically, up to now. While a microscopic dynamical theory for so-called
`simple' liquids is still under construction
\cite{archer06a,archer09,anero07,melchionna08,Marconi09}, the DDFT for
Brownian particles (i.e.\ for colloidal suspensions)
\cite{marconi99,marconi00,archer04b} has been applied with success to a large
number of situations, ranging from phase separation \cite{archer05a} and
spinodal decomposition \cite{archer04a} to the micro-rheology of colloid
polymer mixtures \cite{rauscher07b,gutsche07}\/.

\begin{figure}
\includegraphics[width=\linewidth]{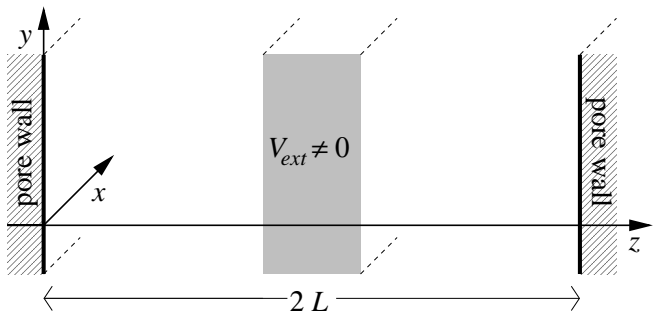}
\caption{\label{setup} We consider a binary colloidal mixture
confined in a three-dimensional slit pore of width $2\,L$\/. In the
center of the slit pore an attractive or repulsive external potential
$V_{{ext}}$ acts on the particles. This external potential can lead
to selectivity of one species of particle over the other in this region
of the slit.
}
\end{figure}

Here, we use DDFT to study the dynamics of selectivity in a binary colloidal
mixture of particles of two different sizes. We consider the situation where
the mixture is confined within a three-dimensional slit pore, where, apart from the usual fluid
ordering near the walls of the slit, the equilibrium fluid is homogeneously
distributed within the slit. At time $t=0$, we switch on an external
potential in the center of the channel (see Fig.~\ref{setup}) and then follow the time evolution of
the system using DDFT\/. We select this configuration since it may be realized
experimentally in a straightforward manner, by using laser tweezers to create 
the external potentials.

This paper is laid out as follows: In Sec.~\ref{DFTtheory} we give a short
overview of static and dynamical DFT and describe the model system under
consideration. Sec.~\ref{resultssec} begins with a discussion of the influence
of external potentials on the equilibrium fluid density distributions. This is
followed by a discussion of the dynamics of selectivity for a binary mixture
of hard spheres. We close with a summary and an outlook in
Sec.~\ref{outlooksec}\/.

\section{Theory and model system}
\label{DFTtheory}

The theories that we use for describing the fluid are static (equilibrium) DFT
and DDFT. Here, we only introduce the aspects of the theories that are
relevant to the present study. For a detailed introduction to DFT, see e.g.\
Refs.\ \cite{hansen06,evans92,evans79} and for an introduction to DDFT,
see Refs.\ \cite{marconi99,marconi00,archer04b,archer04a}\/.  

\subsection{Equilibrium DFT}
\label{equilsec}

We consider a fluid mixture composed of $\nu$ different types of particles 
at a temperature $T$, confined within a fixed volume $\Vol$\/. For systems
where external potentials $\Vext{i}(\vct{r})$ are acting on
particles of type $i$, with $i=1\dots\nu$, treating the system in the grand
canonical ensemble, one can rigorously prove the existence of the grand
potential (free energy) functional $\Omega[\{\rho_i\}]$. The set of equilibrium
fluid density profiles $\{\rho_i^{(\text{eq})}(\vct{r})\}$, are those which
minimize $\Omega[\{\rho_i\}]$\/. The minimum value of $\Omega[\{\rho_i\}]$ is
the grand potential of the system \cite{evans79}. The grand potential
functional may be written in the following way: 
\begin{equation}
\label{grandfunctional}
\Omega[\{\rho_i\}]=\mathcal{F}[\{\rho_i\}] 
+\sum\limits_{i=1}^\nu \int\limits_\Vol
\rho_i(\vct{r})\left[\Vext{i}(\vct{r}) - \mu_i\right]d^3r,
\end{equation}
where $\mu_i$ is the chemical potential for particles of type $i$ and the
intrinsic Helmholtz free energy functional 
\begin{equation}
\label{eq:helmholtz}
\mathcal{F}[\{\rho_i\}]=\sum\limits_{i=1}^\nu 
\frac{1}{\beta}\,\int\limits_\Vol 
 \rho_i(\vct{r})\left[\ln \lambda_i^3\,\rho_i(\vct{r})-1\right] d^3r 
+
\mathcal{F}_{{ex}}[\{\rho_i\}],
\end{equation}
which is the sum of the ideal-gas parts, where $1/\beta= k_B\,T$ is the thermal
energy, $\lambda_{i}$ is the thermal De Broglie wavelength, and
$\mathcal{F}_{{ex}}[\{\rho_i\}]$ is the free energy contribution
originating from the interactions between the particles.  

From the minimization principle on $\Omega[\{\rho_i\}]$, it follows that the 
equilibrium fluid density profiles are the solution of the following
Euler-Lagrange equations: 
\begin{equation} \label{elg}
  \frac{\delta \Omega[\{\rho_i\}]}{\delta \rho_i(\vct{r})} = 0 = 
\beta^{-1} \ln \lambda_i^3 \rho_i(\vct{r}) +
\frac{\delta \mathcal{F}_{{ex}}[\{\rho_i\}]}{\delta  \rho_i(\vct{r})}+
\Vext{i}(\vct{r})-\mu_i.
\end{equation}
For a given set of external potentials $\{\Vext{i}(\vct{r})\}$, these 
yield a set of non-linear equations for the density profiles 
$\{\rho_i(\vct{r})\}$, which can be solved e.g.\ using an iterative numerical
algorithm.

The excess free energy functional $\mathcal{F}_{{ex}}[\{\rho_i\}]$ is
only known exactly for a few one-dimensional model systems. However, for many
experimentally relevant systems, approximate functionals have been constructed
which yield results that are in quantitative agreement with experiments and
simulations. In particular, this is the case for mixtures of hard spheres, for
which functionals arising from fundamental measure theory (FMT), mainly based
on geometric considerations, have been developed
\cite{rosenfeld89,roth02a,hansengoos06}\/. In FMT the excess free
energy functional is given by 
\begin{equation} \label{fmt}
\beta {\cal F}_{ex}[\{\rho_i\}] = \int \Phi(\{n_\alpha(\vct{r})\})\,d^3r\,,
\end{equation}
where the excess free energy density $\Phi$ is a function of a set of weighted
densities 
\begin{equation} \label{weighted}
n_\alpha(\vct{r}) = \sum_{i=1}^\nu \int \rho_i(\vct{r}')\,
\omega_\alpha^{i}(\vct{r}-\vct{r}')\,d^3 r'.
\end{equation}
In Eq.~(\ref{weighted}) the geometrical weight functions are denoted by
$\omega_\alpha^{i}(\vct{r})$ and $\alpha$ labels four scalar and two vector-like
weight functions. It is the fundamental measure theory constructed in Ref.\
\cite{roth02a} that we implement in this study.  

\subsection{Dynamical DFT}

In a manner analogous to the Runge-Gross theorem for quantum systems \cite{runge84}, one
can prove the existence of a dynamical density functional theory for classical 
systems \cite{chan05}\/. As in the static case, the proof of the
existence of a DDFT does not lead to {\em having} a theory with which one can
do calculations. However, for systems of Brownian particles with over-damped
stochastic equations of motion [see Eq.\ \eqref{brownianeq}, below], 
a successful DDFT has been developed by Marconi and Tarazona
\cite{marconi99,marconi00} for calculating the time evolution of the ensemble 
average density profiles $\rho_i(\vct{r},t)$, $i=1\dots \nu$. Note that by
`ensemble average' density, we mean an average over the ensemble of different
realizations of the stochastic noise terms and particle starting positions in
the equations of motion. For a detailed discussion of the differences between
the dynamics of the ensemble averaged, the instantaneous, and the coarse
grained density, see Ref.\ \cite{archer04b}. 

In real three-dimensional systems, the number of particles in a volume $\Vol$
can be changed only by a flux through the boundaries $\partial\Vol$, which, in
absence of chemical reactions, directly leads to local particle conservation. 
In Sec.~\ref{equilsec} above, we discussed DFT, an equilibrium statistical 
mechanical theory for non-uniform fluid mixtures in the grand canonical
ensemble. However, for systems with impermeable boundaries, the equilibrium 
limit leads to the canonical ensemble. For this reason we consider fixed
numbers $N_i$ 
of particles of type $i$ at positions $\vct{r}^{(i)}_n(t)$, 
where ${n=1\dots N_i}$, and we assume that the particle dynamics is governed
by the following over-damped Brownian equations of motion: 
\begin{equation}
\label{brownianeq}
\frac{d\vct{r}^{(i)}_n}{dt} = \Mob_i\,\left[
-\grad \Vext{i}(\vct{r}^{(i)}_n)
+\vct{F}_n^{(i)}
\right]+\sqrt{k_B\,T\,\Mob_i}\,\vct{N}_n^{(i)}(t),
\end{equation}
where $\vct{F}_n^{(i)}$ is the net force on particle $n$ of type $i$, due to the
interactions with all the other particles in the system and $\Mob_i$ is the
mobility coefficient for particles of type $i$\/. $\vct{N}_n^{(i)}$ is a
Gaussian random white noise, with zero mean and the correlator 
\begin{equation}
\left\langle
\vct{N}_{n}^{(i)}(t)\circ \vct{N}_{m}^{(j)}(t')
\right\rangle = 2\,\delta_{nm}\,\delta_{ij}\,
\delta(t-t')\,\bm{1}.
\end{equation}

The Fokker-Planck equation corresponding to Eq.~\eqref{brownianeq} describes
the time evolution of $W(\{\vct{r}_n^{(i)}\}, t)$, the probability of finding
the particles in the system at the positions $\vct{r}_n^{(i)}$
at time $t$\/. By integrating over $W(\{\vct{r}_n^{(i)}\}, t)$ with respect to
all but one of the coordinates of particles of type $i$, one obtains a time
evolution equation for the ensemble averaged density $\rho_i(\vct{r},t)$
\cite{archer05a}. For systems of particles interacting via pairwise
interaction potentials, the resulting time evolution equations for
$\rho_i(\vct{r},t)$ depend only on the non-equilibrium two-body distribution
functions, but for systems interacting via many-body interaction potentials, 
the time evolution equations for $\rho_i(\vct{r},t)$ depend on the higher-body
distribution functions \cite{archer04a}\/. The time evolution equations for
$\rho_i(\vct{r},t)$ are the lowest members in a hierarchy of $\sum_{i=1}^\nu
N_i$ equations that is similar to the Bogoliubov-Born-Green-Kirkwood-Yvon
(BBGKY) hierarchy \cite{archer04b,hansen06, archer05a, archer06a, archer09}. 
The DDFT is obtained by using a closure relation to truncate the hierarchy.
This is done by approximating the non-equilibrium two-body and higher-body
distribution functions by the same quantities in an equilibrium system with
the same one-body density distributions
\cite{marconi99,marconi00,archer04a}. One can show that for a system with a
given set of density distributions $\{ \rho_i(\vct{r})\}$, one can find a
unique set of external potentials $\{ \Vext{i}(\vct{r})\}$, such that the
system would be at equilibrium if it was exposed to these external potentials. 
This closure leads to the following set of coupled DDFT equations
\cite{archer05a}: 
\begin{equation}
\label{ddfteq}
\frac{\partial \rho_i(\vct{r},t)}{\partial t} = 
\Mob_i\,\grad\cdot \left[
\rho_i(\vct{r},t)\,\grad\frac{\delta
\Omega[\{\rho_i\}]}{\delta\rho_i}\right],
\end{equation}
where the functional $\Omega[\{\rho_i\}]$ is the {\em equilibrium} grand 
potential functional, given in Eq.~\eqref{grandfunctional}\/. Note, that the
gradient of the variation of the grand potential equals the gradient of the
variation of the total free energy, i.e., the sum of the intrinsic
free energy $\mathcal{F}[\{\rho_i\}]$ and the contributions from the
external potentials $V_{{ext}}$\/. For non-interacting particles,
Eq.~\eqref{ddfteq} reduces to the drift-diffusion equation. The (equilibrium)
closure approximation used to obtain these equations means that there are some
limitations on what kinds of problems the DDFT can be applied to. For example,
one can not use it to describe barrier crossing in systems with free energy
barriers. However, some aspects of glassy systems can be described within the
DDFT framework \cite{archer07a}\/.

In order to implement Eq.~(\ref{ddfteq}) using the FMT approximation for
$\Omega[\{\rho_i\}]$, we may take advantage of certain simplifications that
arise due to the structure of the FMT when calculating the gradient of the
variation of the excess free energy. We present this in the Appendix.

\subsection{Model system}

The model system that we study is a binary mixture of hard-spheres, composed
of small ($s$) and big ($b$) particles with sphere radii $R_s$ and $R_b$,
respectively. The particle densities are $\rho_s(\vct{r},t)$ and
$\rho_b(\vct{r},t)$\/. We first consider a
bulk mixture with given densities and examine the influence of external
potentials, which can be either attractive or repulsive, that act on a
localized region. 

Depending on the origin of the external potentials, they may
act in the same way on both types of particles, or alternatively the potentials
may be proportional to the
volume of the particles. In studies of ion channels with wide selectivity
filters with fixed charges, such as the l-type calcium channel, it was found
that a competition between energy and entropy can explain size selectivity
\cite{Nonner00,Nonner01,Gillespie02}. In such ion-channels the external
potentials acting on the ions are generated by the electric field due to the
charges fixed in the selectivity filter. Since the potentials generated in
this way are the same for particles carrying the same charge, we model these
by setting the potentials acting on both species of particles to be equal,
i.e.\ we set $\Vext{s}(\vct{r})=\Vext{b}(\vct{r})$, to model a mixture with
two sizes of particles having the same sign and magnitude charge on them.

However, one may also consider a different situation: that of e.g.\ optical
tweezers applying a force on colloidal particles within a specified region of
space. When the forces acting on a colloidal mixture are generated by optical
tweezers, then the external potentials acting on the particles are
proportional to the volume of the particles. To model this situation,
one should set $\Vext{s}(\vct{r})= (R_s/R_b)^3 \Vext{b}(\vct{r})$. 

\begin{figure}
\includegraphics[width=0.8\columnwidth,clip]{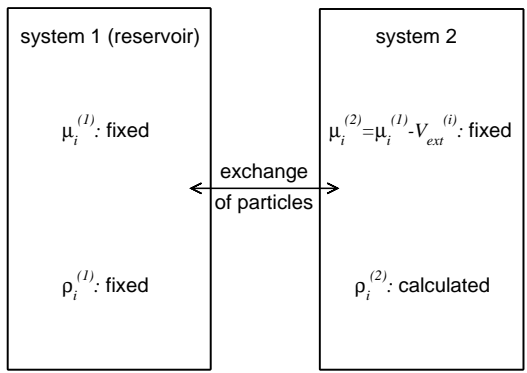}
\caption{\label{fig:system} The effect of size selectivity can be rationalized
  in a simple model of two coupled systems. In system 1, which acts as a
  reservoir for system 2, the densities of the mixture are fixed to be
  $\rho_i^{(1)}$. In system 2, the chemical potentials are shifted relative to
  the corresponding values in system 1 by the external potentials
  $V_{ext}^{(i)}$, which are considered to be constant throughout the
  system. The different particle densities in system 2 result from these shift
  in the chemical potentials.}
\end{figure}

The mechanism of the size selectivity as studied in the next section can be understood
by considering a simplified system 
of a binary colloidal fluid mixture partitioned into two parts:
in the first part, a finite region of space, in which the external potentials
are applied, and the second part, which acts as a `reservoir', consisting of
the remaining parts of the system. If we assume both systems to be infinite in
size (i.e., we neglect interface and wall effects) and if we assume
$V_{ext}^{(i)}$ to be constant in system~2, we arrive at the situation
depicted in Fig.~\ref{fig:system} with the chemical potentials $\mu_i^{(1)}$
and $\mu_i^{(2)}$ in system~1 and 2, respectively given by
\begin{equation}
\mu_i^{(\ell)}=\mu_i^{(\ell)}(\{\rho_j^{(\ell)}\}) = \left.\frac{\delta
\mathcal{F}}{\delta \rho_\ell}\right|_{\left\{\rho_j=\rho_j^{(\ell)}\right\}}
\end{equation}
for $i=s,b$ and $\ell=1,2$ and with the intrinsic free energy
$\mathcal{F}[\{\rho_i\}]$ defined in Eq.~\eqref{eq:helmholtz}\/.
$\rho_j^{(\ell)}$ is the spatially constant density of species $j$
in system $\ell$\/.

In system 1 (which is the reservoir) we 
specify the densities of the mixture, $\rho_i^{(1)}$, where
$i=s,b$, which fixes the chemical potentials in system~1\/.
In system~2, which is coupled to system 1, so that particle exchange with the
reservoir is allowed, we apply the (spatially constant) external potentials
$V_{ext}^{(i)}$\/.  Applying these external potentials is then
equivalent to shifts in the chemical potentials in
system~2\/. The densities in system~2, $\rho_i^{(2)}$, can thus be computed by
solving the following set of equations 
\begin{equation} \label{mu}
\mu_i^{(1)}(\{\rho_i^{(1)}\}) = \mu_i^{(2)}(\{\rho_i^{(2)}\}) +
V_{ext}^{(i)}\quad\text{for}\quad i=s,b.
\end{equation}
If the external potentials are attractive, then the chemical potentials in
system~2 are effectively increased over those in system 1, while they are
decreased in the case of repulsive potentials. In the low density limit, the 
Eqs.~(\ref{mu}) for $i=s$ and $i=b$ decouple into a pair of independent
equations. At higher densities, Eqs.~(\ref{mu}) form a non-linear pair of
coupled equations for the densities in system~2\/. As discussed in
Sec.~\ref{outlooksec} below, these equations allow us to understand the
phenomena described in the next section\/.

\section{Results}
\label{resultssec}

\subsection{The external potentials}
\label{potsec}
\begin{figure}
\includegraphics[width=1.\columnwidth]{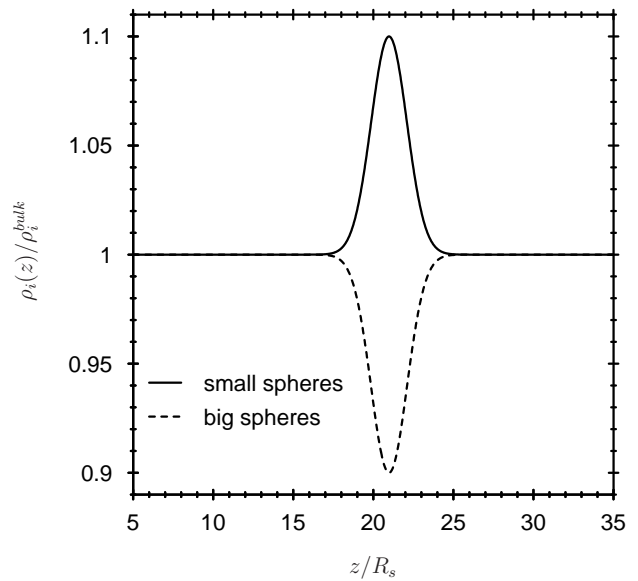}
\caption{\label{fig:profs} Density profiles $\rho_i(z)$, where $i=s,b$,
  divided by the respective bulk densities $\rho_i^{bulk}$
  (corresponding to the bulk packing fraction
  $\eta_i=\frac{4}{3}\,R_i^3\,\pi\,\rho_i^{bulk}$),
  of a binary mixture of hard spheres with a size
  ratio of $R_b=2 R_s$. The density profiles show small-particle
  selectivity: locally, with the help of the external potentials, the density
  of the small spheres (solid line) is slightly increased, while the density
  of the big spheres (dashed line) is slightly decreased. Using
  Eq.~(\ref{vext}), it is possible to calculate the external potentials
  $V_{ext}^i(z)$, $i=s,b$, which give rise to these equilibrium density
  profiles. These external potentials are displayed in Fig.~\ref{fig:pot_mix},
   for various bulk packing fractions.}
\end{figure}

\begin{figure}
\includegraphics[width=1.\columnwidth]{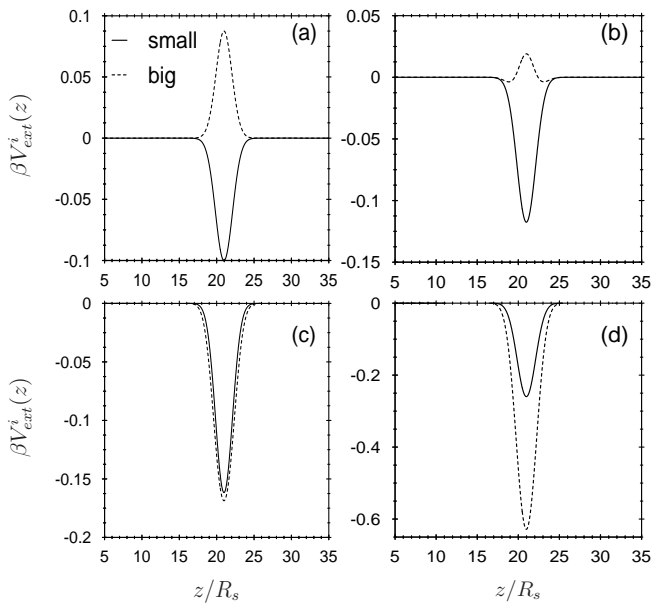}
\caption{\label{fig:pot_mix} In order to generate a slight local increase in 
  the density of the small spheres and a slight decrease in the density of the
  big spheres (as displayed in Fig.~\ref{fig:profs}), it is necessary to apply
  external potentials $V_{ext}^i(z)$. Four examples are displayed here for a
  binary mixture with size ratio $R_b=2 R_s$ and equal bulk packing
  fractions $\eta_s=\eta_b$. At low bulk packing fractions $\eta_s=\eta_b=0.01$, we
  see in (a) that an attractive potential has to act on the small spheres and
  a repulsive potential on the big spheres, in order to generate local
  small-particle selectivity. This can be understood from the ideal-gas
  limit. In (b), where $\eta_s=\eta_b=0.04$, the repulsion acting on the big
  spheres is significantly reduced, as a result of the competition between
  entropy and energy. In (c), where $\eta_s=\eta_b=0.09$, both external
  potentials are roughly the same and are attractive and in (d), where
  $\eta_s=\eta_b=0.15$, the attraction required to attract the small particles
  is weaker than the {\em attraction} required to effectively {\em repel} the
  big particles.}
\end{figure}

Within the framework of static DFT, the equilibrium thermodynamic properties
of the system are obtained from
the solution of the Euler-Lagrange equations \eqref{elg},
obtained by minimizing the functional in Eq.~\eqref{grandfunctional}\/.
Typically, one specifies the external potentials
$\Vext{i}(\vct{r})$ and then computes the resulting equilibrium density
distributions $\rho_i(\vct{r})$, $i=1,\dots,\nu$. 
However, it is also possible to employ the Euler-Lagrange equations
\eqref{elg} in order to
compute the external potentials that give rise to a {\em specified} set of
density distributions $\{\rho_i(\vct{r})\}$ \cite{henderson91}. One obtains:
\begin{equation} \label{vext}
\Vext{i}(\vct{r})=-\frac{\delta \mathcal{F}_{{ex}}[\{\rho_i\}]}
{\delta \rho_i(\vct{r})}-\beta^{-1}
\ln \lambda_i^3 \rho_i(\vct{r}) + \mu_i.
\end{equation}
It is worth noting that in the limit of vanishing densities, the
contribution due to particle interactions, which depends on the variation of
$ \mathcal{F}_{{ex}}[\{\rho_i\}]$, becomes negligible and one recovers
the ideal-gas result: $\Vext{i}(\vct{r})=-\beta^{-1} \ln \lambda_i^3
\rho_i(\vct{r}) + \mu_i$. In the ideal-gas limit, where the density
distributions are simply the Boltzmann factors of the external potentials, it
is clear that in order to locally increase the density of the particles over
the bulk value, attractive potentials are required, and to locally decrease
the density, repulsive potentials are needed. 

For a one-component interacting system, the behavior is similar to the
ideal-gas case, although the density distribution is not simply the Boltzmann
factor of the external potential. The density can locally be increased
or decreased by an attractive or repulsive external potential, respectively.

The situation becomes more interesting in the case of a binary 
mixture. If we start with constant densities in an infinite bulk
system and wish to generate a set
of density profiles such as those shown in Fig.~\ref{fig:profs}, which show a
local increase in the density of the small spheres and at the same location a
decrease in the density of the big spheres, we can use Eq.~(\ref{vext}) to
calculate the external potentials required to achieve this. 

In Fig.~\ref{fig:pot_mix} we display the external potentials that one must
exert on the system in order to observe  the two density profiles
shown in Fig.~\ref{fig:profs}, for a binary hard-sphere mixture with
$R_b=2 R_s$ and equal bulk packing fractions 
of the small and large particles, $\eta_s$
and $\eta_b$, respectively
($\eta_i=\frac{4}{3}\,R_i^3\,\pi\,\rho_i^{bulk}$, where
$\rho_i^{bulk}$ is the density of species $i$ in the bulk, at points in space a
large distance from the support of the external potential)\/.
At low values of $\eta_s=\eta_b=0.01$, we see in
Fig.~\ref{fig:pot_mix}(a), that the required external potentials are similar
to those predicted by the ideal-gas functional. The external potential that
gives rise to a local increase in the density of the small spheres is
attractive (solid line) and the external potential that gives rise to a
decreased in the density of the big spheres is repulsive (dashed
line). However, if the bulk packing fractions increase to
$\eta_s=\eta_b=0.04$, as shown in Fig.~\ref{fig:pot_mix}(b), we find that the 
external potential acting on the big particles is 
significantly less repulsive than in the low density case
displayed in Fig.~\ref{fig:pot_mix}(a). This 
stems from the competition between energy, due to the interactions of the
particles with the external potentials, and entropy, due to the inter-particle
interactions and the resulting excluded volume. The influence of the
competition between energy and entropy is  more pronounced
if we increase the bulk packing fractions even
further. For $\eta_s=\eta_b=0.09$, in Fig.~\ref{fig:pot_mix}(c) we see that
both species of particles have to be subjected to an
attractive potential well, in order to
obtain density profiles that locally have an increase in the density of
the small particles, at the same time as a decrease
in the density of the big particles. For this
particular choice of packing fractions, the external potentials for both
sizes of particles are roughly the same. In order keep the 
reduction of the density of the large particles at the same
moderate level at much
higher packing fractions $\eta_s=\eta_b=0.15$, the attraction on
the big particles has to be much
stronger than the attraction on the small particles.

\subsection{Equilibrium DFT for selectivity}
\label{eqselectsec}
\begin{figure}
\includegraphics[height=0.5\columnwidth]{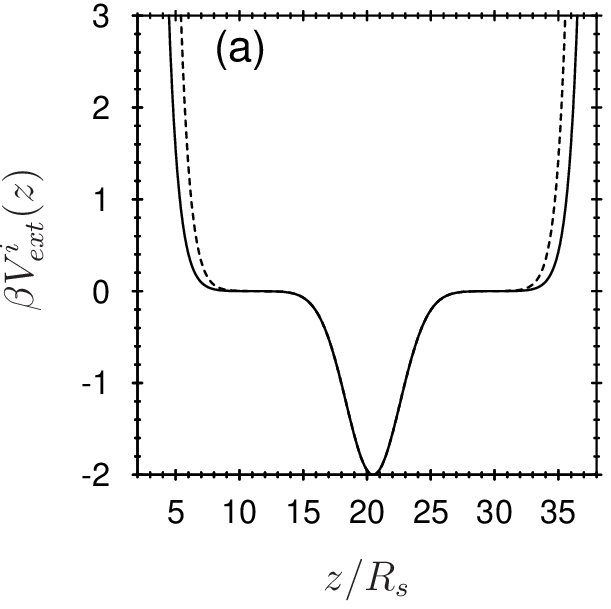}~\includegraphics[height=0.5\columnwidth]{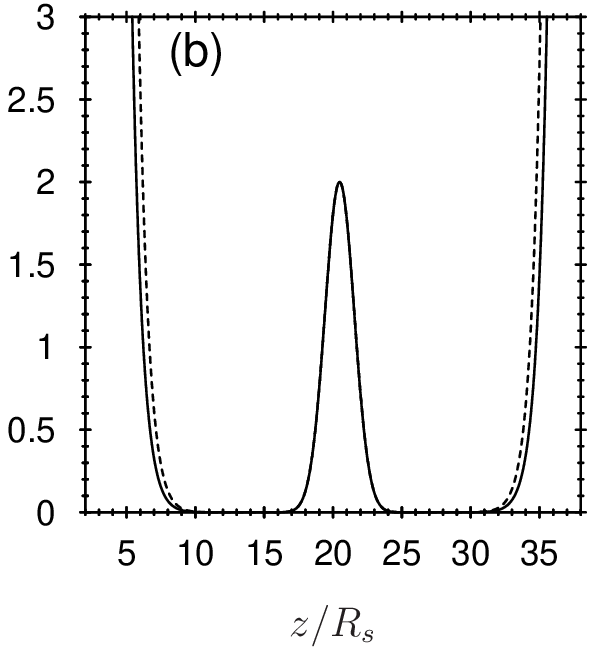}
\caption{\label{fig:pot_small_big} The external potential acting on the small
  spheres (solid lines) and big spheres (dashed lines). In addition to the soft slit
  potentials, we perturb the density distributions by additional potentials
  in the middle of the slit, which can be either (a) attractive or (b)
  repulsive.} 
\end{figure}

\begin{figure}
\includegraphics[height=0.48\columnwidth]{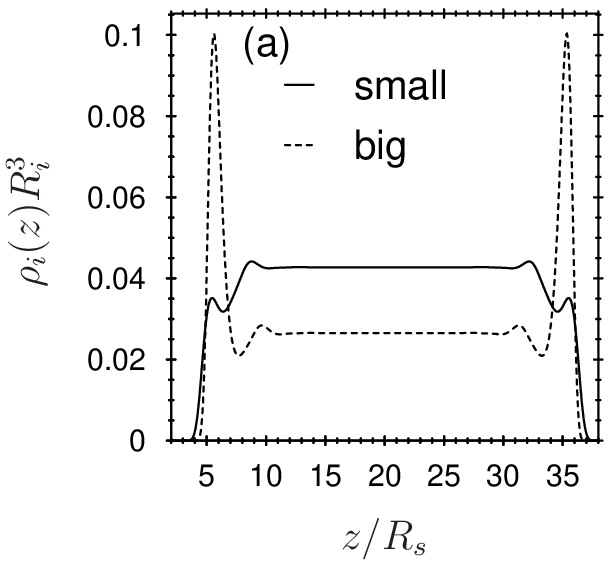}~\includegraphics[height=0.48\columnwidth]{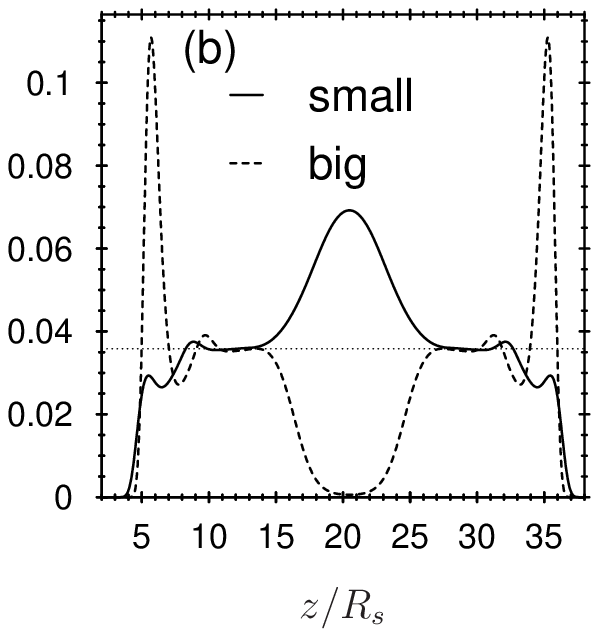} 
\caption{\label{fig:small} 
(a) The equilibrium density profiles of a binary
  hard-sphere mixture with $R_b=2 \,R_s$
  confined within a
  slit with potentials given by Eq.~\eqref{eq:extpot1}\/. (b) When
  the attractive potentials in the middle of the slit are switched
  on [see Fig.~\ref{fig:pot_small_big}(a) and
  Eq.~\eqref{eq:extpot2}], the density of the small spheres is locally enhanced
  in the middle of the slit while the density of the big spheres
  is decreased. This is an example of small particle selectivity.
  In both (a) and (b) the chemical potentials $\mu_b$ and $\mu_s$ are
  chosen such that the adsorptions given by Eq.~\eqref{eq:ads}
  are $\Upsilon_b R_b^2=0.495$ and
  $\Upsilon_s R_s^2=1.308$\/. The chemical potentials in (b) correspond
  to reservoir packing fractions $\eta_a = \eta_s = 0.15$\/, as indicated by
  the dotted line.}
\end{figure}

\begin{figure}
\includegraphics[height=0.48\columnwidth]{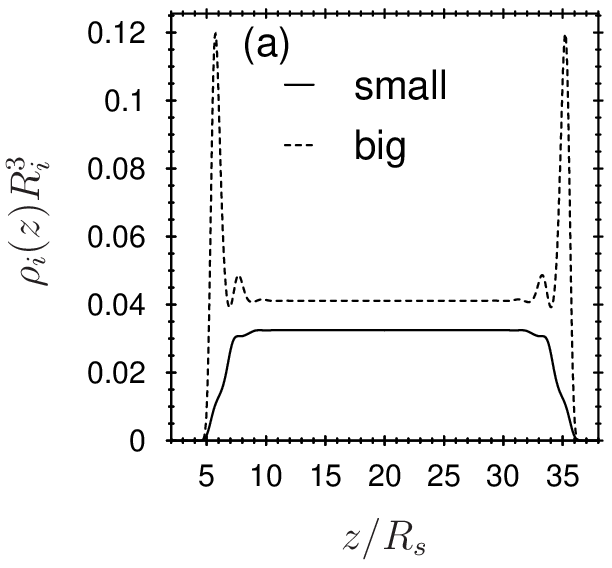}~\includegraphics[height=0.48\columnwidth]{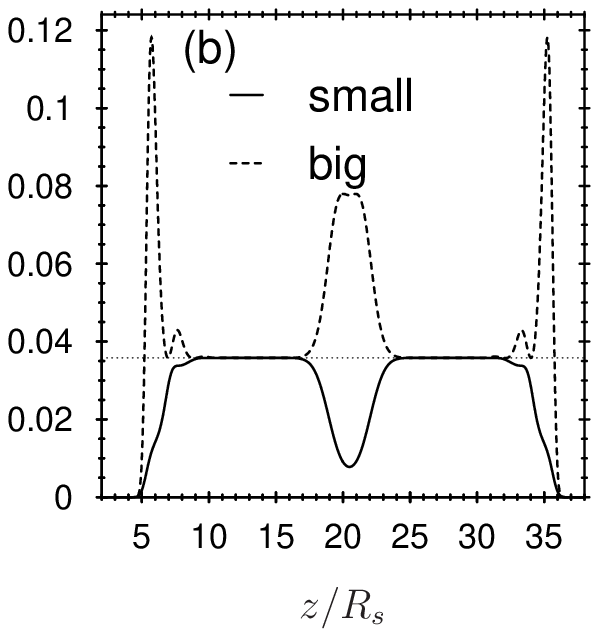}
\caption{\label{fig:big} 
Same as Fig.~\ref{fig:small}, but  for repulsive potentials.
(a) The equilibrium density profiles with $R_b=2 \,R_s$
  confined within a
  slit with potentials given by Eq.~\eqref{eq:extpot1}\/. (b) When
  the repulsive potentials in the middle of the slit are switched
  on [see Fig.~\ref{fig:pot_small_big}(b) and
  Eq.~\eqref{eq:extpot2}], the density of the small spheres is
  locally reduced
  in the middle of the slit while the density of the big spheres
  is increased. This is an example of large particle selectivity.
  In both (a) and (b) the adsorptions  given by Eq.~\eqref{eq:ads}
  are $\Upsilon_b R_b^2=0.786$ and
  $\Upsilon_s R_s^2= 0.905$\/.  The chemical potentials in (b) correspond
  to bulk reservoir fractions $\eta_a = \eta_s = 0.15$\/, as indicated by the
  dotted line.
  }
\end{figure}

From the results presented so far, we can see that the competition between
energy and entropy in mixtures can lead to surprising effects. The question
that arises is whether or not it is possible to use this competition in order
to select one species over the other with the aid of an appropriately chosen
external potential. To this end, we consider a binary hard-sphere mixture
confined between two planar walls -- i.e.\ in a slit. We model the potentials
due to the walls of the slit as follows: 
\begin{equation}
\label{eq:extpot1}
\Vext{i}(z)=A\left( \frac{z-z_0}{L_i}\right)^m,
\end{equation}
where $A=10\,k_B\,T$, $z_0=20.48\, R_s$, $m=20$ and $L_i=L+R_i$, where
$L=15\, R_s$. Due to the large value of the power $m$, Eq.~(\ref{eq:extpot1})
models the continuous potentials of a pair of slightly soft parallel planar
walls, separated a distance $2\,L$. $z_0$ is the center of the slit
pore. When the fluid mixture in the slit is at
equilibrium, we perturb the fluid by introducing additional external
potentials in the middle of the slit, so that the external potentials are:
\begin{equation}
\label{eq:extpot2}
\Vext{i}(z)=A\left( \frac{z-z_0}{L_i}\right)^m
+\epsilon_i\exp\left[ \left( \frac{z-z_0}{w} \right)^2\right].
\end{equation}
The additional second term in the external potentials can be either
attractive ($\epsilon_i<0$), as shown in Fig.~\ref{fig:pot_small_big}(a), or
repulsive ($\epsilon_i>0$), as
indicated in Fig.~\ref{fig:pot_small_big}(b). The range of the external
potential is $w=3\,R_s$ and the depth or height
$\epsilon_i=\pm 2 \,k_B\, T$\/. 

First, we consider the case of an attractive potential in the middle of the
slit ($\epsilon_i<0$). We compare two equilibrium states: (i) the binary
mixture in the slit without the additional potentials ($\epsilon_i=0$), and
(ii) the equilibrium state of the binary mixture with the potentials
($\epsilon_i=-2k_BT$), shown in Fig.~\ref{fig:pot_small_big}(a). We calculate
the density profiles which minimize the equilibrium DFT by solving the
Euler-Lagrange equations, Eq.~(\ref{elg}). Since equilibrium DFT is defined in
the grand canonical ensemble, particle exchange between the system
and a
reservoir, which sets the chemical potentials, is possible. However, we wish
to compare our equilibrium DFT calculations with DDFT computations, which
preserves the number of particles. In order to make an honest comparison, we
minimize the equilibrium DFT under the constraint that the amount of material
within the slit is fixed, i.e., that the adsorptions
\begin{equation}
\Upsilon_i=\int_{-\infty}^{\infty}\rho_i(z) dz,
\label{eq:ads}
\end{equation}
are fixed. To enforce these constraints, we treat the chemical
potentials $\mu_a$ and $\mu_b$ as
Lagrange multipliers and we choose them such that in the case
when the potentials in the middle of the slit are switched on, the reservoir
packing fractions are $\eta_s=\eta_b=0.15$\/. In the case of an
attractive potential this results in $\Upsilon_s R_s^2=1.308$ and
$\Upsilon_b R_b^2=0.495$, while in the repulsive case the corresponding
adsorptions are $\Upsilon_s R_s^2=0.905$ and $\Upsilon_b R_b^2=0.786$\/.
Correspondingly, the reservoir packing fractions when the
potentials are switched off are different.

When the attractive potential wells are not switched on, we find that the
density profile of the big spheres close to the walls of the slit possess
prominent peaks [see Fig.~\ref{fig:small}(a)], indicating
strong packing and correlation effects at the walls of the slit. The density
of the small spheres is slightly decreased at the wall. In the middle of the
slit we find a region where both densities are constant. When the attractive
potentials in the center of the slit are switched on, so that {\em both} the
small and the big particles are subjected to the external potentials shown in 
Fig.~\ref{fig:pot_small_big}(a), we find that there is a significant
local increase in the density of the small particles in the region of the
attraction, while the density of the big particles is strongly decreased here
-- see Fig.~\ref{fig:small}(b). Hence, if the bulk densities of the big and
the small particles are sufficiently large, an attractive potential well can
be small particle selective due to the competition between energy and
entropy.

We now turn our attention to the case when a repulsive potential is turned on
in the middle of the slit -- see Fig.~\ref{fig:pot_small_big}(b). In
Fig.~\ref{fig:big}(a) we display the equilibrium density profiles for the case
when the repulsive potentials are not switched on. The profiles are similar to
those shown in Fig.~\ref{fig:small}(a)\/. The difference is due to our treatment
of the chemical potentials as Lagrange multipliers to set the adsorptions in
Eq.~\eqref{eq:ads}, for the case when the potentials in the middle of the
slit are switched on. This leads to the adsorptions and therefore
also to the density profiles being different in
the two cases when the potentials are switched off. 

In Fig.~\ref{fig:big}(b) we display the equilibrium fluid density profiles 
for the case when the repulsive potentials in the middle of the slit, for both
components, are switched on. In the region where the repulsive external
potentials are applied, the local density of the small spheres is decreased
and the density of the big spheres is {\em increased}. It is interesting to
note that despite the repulsion acting on the big particles, the local density
of the big spheres is greater than the density in the surrounding fluid. This
is an example of big-particle selectivity.

\subsection{Dynamics of selectivity}
\label{dynselectsec}
\begin{figure*}
\includegraphics[width=\textwidth]{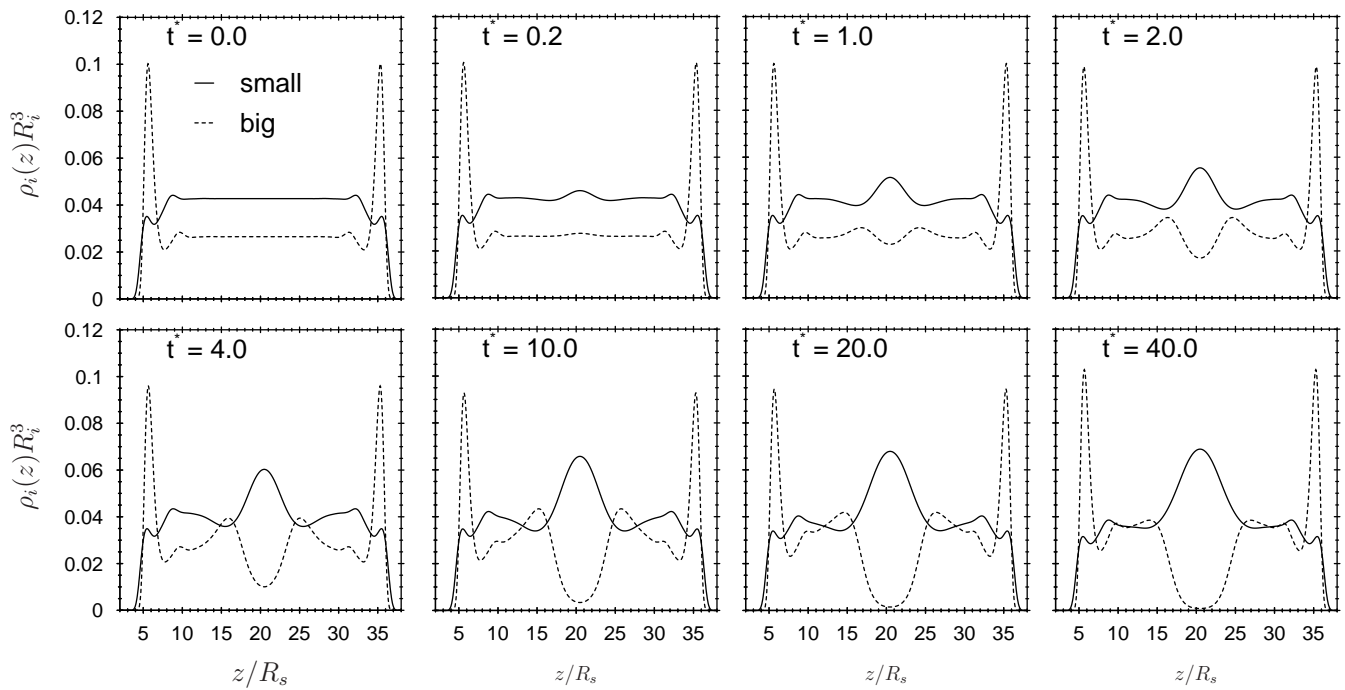}
\caption{\label{fig:small_ddft}  Time evolution  of the density profiles of
  the small (solid lines) and big (dashed lines) spheres from the
  initial state in Fig.~\ref{fig:small}(a) towards the final
  state in Fig.~\ref{fig:small}(b)\/. At time $t^*=0$ we
  switch on the attractive potentials in the middle of the slit. 
  Note that for short times, both species of particles follow the attraction
  and move towards the center. However, for larger
  times $t^*>1$, only the density of the small spheres increases in the region
  of the potential well, while the big spheres move away from the center,
  despite the attractive external potential.}
\end{figure*}

\begin{figure*}
\includegraphics[width=\textwidth]{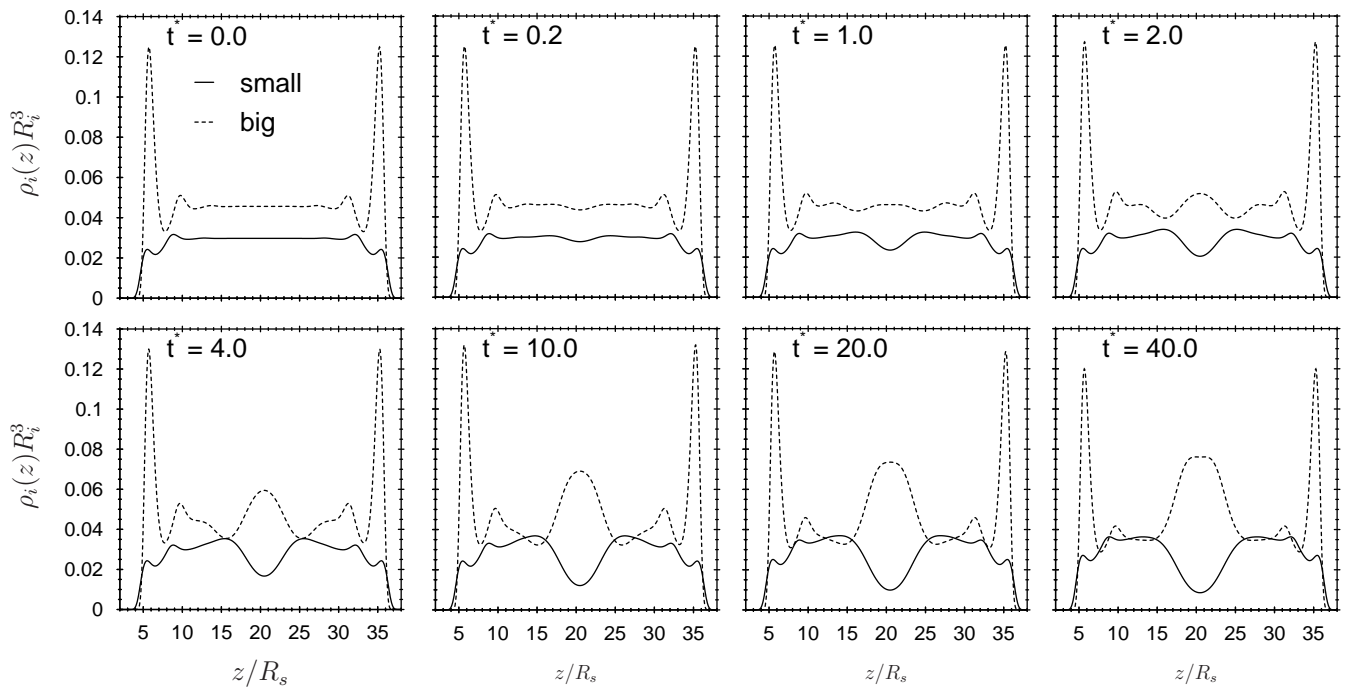}
\caption{\label{fig:big_ddft}  Time evolution of the density profiles of the
  small (solid lines) and big (dashed lines) spheres from the
  initial state in Fig.~\ref{fig:big}(a) towards the final
  state in in Fig.~\ref{fig:big}(b)\/. At time $t^*=0$ we switch
  on the repulsive potentials in the center of the slit. 
  Note that for short times, both species of particles follow the repulsion
  and move away from the center. However, for larger
  times $t^*>1$, only the density of the small spheres decreases in central
  region, while the density of the big spheres increases in the center,
  despite the repulsive external potential.}
\end{figure*}

In order to better understand the behavior of the system, we study the time
evolution of the density profiles of both types of particles as they evolve
between the two equilibrium states shown in Fig.~\ref{fig:small}, the case
with the attractive potentials applied in the center of the slit, and those
shown in Fig.~\ref{fig:big}, the case where the repulsive potentials are
exerted in the center of the slit. In what follows, all times
$t^*\equiv t/\tau$ are given in units of the Brownian time-scale 
$\tau=(\beta/\gamma) R_s^2$; it is roughly the time it takes for a particle to
diffuse over a distance equal to its own radius. Note also that we have set
the two mobility coefficients to be equal,
$\gamma_1=\gamma_2=\gamma$\/. In a colloidal suspension this would not be the
case: the mobility of spherical particles in a suspension decreases with the
inverse of the cross-sectional area. For the systems under consideration here
this results in $\gamma_s/\gamma_b = 4$\/. With the small particles being
considerably faster than the big ones, the initial behavior of the big
particles described in the following might not be observable in colloidal
suspension.

The initial conditions for attractive
and repulsive potentials in the center of the slit pore are those
displayed in Figs.~\ref{fig:small}(a) and \ref{fig:big}(a),
respectively.

In the first case, when we switch on the attractive potentials in the center
of the slit at time $t^*=0$, initially both types of particles behave as one
would intuitively expect: they follow the attraction and move towards the 
center of the slit. However, this is only the case for short times. This
initial influx of particles results in a slight increase in the particle
densities at the center of the slit, as can be seen in the density profiles
shown in Fig.~\ref{fig:small_ddft}. For times $t^*>1$, the drift diffusion
behavior of the system qualitatively changes, as the competition between the
energy due to the interactions of the spheres with the external potentials and
the entropy due to the hard-sphere interactions between particles, sets
in. For times $t^*>1$, only the small particles follow the attraction and the
number density of the small particles in the center of the slit increases
further, while the big spheres are expelled from this region. At short times,
only spheres in the center of the slit show a net movement and the density
profiles do not change over time in the vicinity of the slit walls. However,
at longer times, there is a net flux of the small particles  from the walls of
the slit towards the center and there is also a net flux of big particles from
the center towards the walls. This diffusion process is slow and the time it
takes for the system to reach the final equilibrium state depicted
in Fig.~\ref{fig:small}(b) is rather long. It also depends on the
system size -- the wider the slit, the longer it takes for the particles to
diffuse (say) from the wall to the center. 

We observe similar behavior in the case when we apply repulsive
potentials at the center of the slit. When we switch on the repulsive
potentials at $t^*=0$, both types of particles behave as our intuition
suggests, and they follow the repulsion and move way from the center of the
slit. This initial flow results in a slight decrease in the particle densities
at the center of the slit, as can be seen in the density profiles shown in
Fig.~\ref{fig:big_ddft}\/. However, for longer times $t^*>1$, the behavior of
the system qualitatively changes, as the competition between the energy and
the entropy sets in. For times $t^*>1$, only the small particles follow the
repulsion and the number density of the small particles in the center of the
slit decreases further, while the big spheres become effectively attracted
towards this region, and the density of the big particles increases at the
center of the slit. The final equilibrium state the one shown in
Fig.~\ref{fig:big}(b)\/.

Note that these counterintuitive results occur only when the densities of the
particles are high enough. At low densities, when the system can be modelled
as an ideal-gas, attractive potentials always lead to an increase in the
densities of both types of particles and repulsive potentials always lead to a
decrease. 

\section{Discussion}
\label{outlooksec}

Using the bulk approach depicted in Fig.~\ref{fig:system} and expressed in
Eqs.~(\ref{mu}) it is possible to understand the results presented 
in Sec.~\ref{resultssec}\/. If the external potentials $V_{ext}^{(i)}$ are the
same for all components of the mixture, as assumed in our DFT and DDFT
calculations, we can reproduce the increase or decrease of
the particle densities in the region where the external potentials are applied:
for a binary mixture with size ratio $R_b=2\,R_s$ and bulk packing fractions of
$\eta_s=\eta_b=0.15$ in the reservoir system~1 (corresponding to
$\rho_s^{(1)}\,R_s^3=\rho_b^{(1)}\,R_b^3=0.036$) and an attractive
external potential of magnitude $2\,k_B\,T$,
the density of the small particles in system~2 increases by a factor of 1.91,
while the density of the big particles is reduced to 0.008 of its original
value. This is in good agreement with the results displayed in
Fig.~\ref{fig:small}: the density of the big particles at the
center of the slit pore is negligibly small while the density of the
small particles at the same spot is about twice the reservoir density\/. 
In the case of a repulsive potential of
magnitude $2\,k_B\,T$, the density of the big particles increases
by a factor of 2.25, while the
density of the small spheres is reduced to 0.21 of its value without the
field. Again, this agrees rather well with the results shown in
Fig.~\ref{fig:big}\/. 

While we have shown results for size selectivity only for one size ratio,
i.e. $R_b=2\,R_s$, we have confirmed that a binary mixture with less
asymmetric radii shows analogous behavior. However, in the case of a less
asymmetric size ratio, the amplitude of the attractive or repulsive potential
has to be larger to generate a degree of size selectivity similar to that
reported in our study.

In addition, Eqs.~(\ref{mu}) allows one to easily estimate
the influence of the external potentials in other scenarios and determine at
what densities the system should or should not display selectivity. For
example, if the external potentials are proportional to the volume of the
particles, which would be the case in an experimental realization using laser
tweezers, then one should expect results similar to those reported here, for a
binary mixture with a size ratio $R_b=2\,R_s$ and packing fractions
$\eta_s=0.15$ and $\eta_b = 0.02$\/. In this case, external fields on the
order of  $V_{ext}^{(s)}=\pm 1 k_B T$ and 
$V_{ext}^{(b)}=\pm 8 k_B T$ are sufficient to generate selectivity.

In the present study, the relatively wide slit geometry has only a small effect on the
selectivity observed. However, for more narrow pores, the grand canonical calculations
in a cylindrical channel geometry in Ref.\ \cite{roth05a}\/ showed that the confinement
can enhance the selectivity.

The arguments presented above are based purely on the equilibrium
free energy and hold for
any type of underlying dynamics, i.e., for the overdamped Brownian
dynamics considered here as well as for systems following
Hamiltonian dynamics. A key feature of the dynamics of separation
presented in Figs.~\ref{fig:small_ddft} and \ref{fig:big_ddft} is
that the big particles, although they are finally driven out of the
region in which the attractive potential acts and into the region
in which the repulsive potential acts, initially follow the
direction given by the gradient of the external potential. For
Figs.~\ref{fig:small_ddft} and \ref{fig:big_ddft} we assumed equal
mobilities for the big and the small particles. In a colloidal
suspension this is not be the case (the mobility of spherical
particles in a suspension decreases with the inverse of the
cross-sectional area) and so the initial behaviour of the big particles
described above might not be observable in a colloidal suspension.

To conclude, we should remind the reader that the DDFT in Eq.~\eqref{ddfteq}
that we have used to describe the system, does not incorporate hydrodynamic
interactions between the colloidal particles. Whilst hydrodynamic interactions
do not affect the static properties of the system (i.e.\ the equilibrium DFT
is still applicable), hydrodynamic interactions do have an influence on the
dynamical properties of the system. The influence of hydrodynamic interactions
between the particles has been incorporated in the DDFT in several ways
\cite{royall07,rex09}. We believe that extending the present study to include
the influence of hydrodynamic interactions would not qualitatively change any
of the results that we observe since these only influence the dynamics but not
the energetics of the system. It is the energetics which determine the final equilibrium
state, i.e.\ whether selectivity can be observed or not.

\appendix*
\section{DDFT and the Structure of FMT}

For the calculation of the time evolution of the density profiles, the
gradient of $\delta \Omega/\delta \rho_i({\bf r})$ are required in
Eq.~(\ref{ddfteq}). This includes the gradient of the variation of the excess
free energy functional ${\cal F}_{ex}$. Since we employ FMT
for the excess free energy, we can make use of the the structure of 
${\cal F}_{ex}$, given in Eqs.~(\ref{fmt}) and (\ref{weighted}).
Note that in the general, three-dimensional case, the variation of the excess
free energy with respect to the density profile of component $i$ can be written as
\begin{eqnarray} \label{c1}
\frac{\delta {\cal F}_{ex}}{\delta \rho_i({\bf r})} & = & \sum_\alpha \int
\left. \frac{\partial \Phi}{\partial n_\alpha} \right|_{{\bf r}'}
\frac{\delta n_\alpha({\bf r}')}{\delta \rho_i({\bf r})} ~ d^3 r' \nonumber \\
& = & \sum_\alpha \int
\left. \frac{\partial \Phi}{\partial n_\alpha} \right|_{{\bf r}'}
\omega_\alpha^i({\bf r}'-{\bf r})~ d^3 r'.
\end{eqnarray}
It follows from Eq.~(\ref{c1}) that the gradient acts solely on the weight
functions:
\begin{equation} \label{grad}
\grad \frac{\delta {\cal F}_{ex}}{\delta \rho_i({\bf r})} = \sum_\alpha \int
\left. \frac{\partial \Phi}{\partial n_\alpha} \right|_{{\bf r}'}
\grad \omega_\alpha^i({\bf r}'-{\bf r})~ d^3 r'.
\end{equation}
Although the sum in Eqs.~(\ref{c1}) and (\ref{grad}) is over four scalar
($\alpha=3,\dots,0$) and two vector-like ($\alpha=v2,v1$) weight functions, one
can exploit the relations between the scalar weight functions $4 \pi R_i^2~
\omega_0^i = 4 \pi R_i~ \omega_1^i = \omega_2^i$ and the vector-weighted
functions $4 \pi R_i~ \omega_{v1}^i = \omega_{v2}^i$. This allows one to
reduce to the sum in Eqs.~(\ref{c1}) and (\ref{grad}) to three terms by
introducing the auxiliary functions
\begin{equation}
\Psi_3^{(i)}({\bf r})=\frac{\partial \Phi}{\partial n_3},
\end{equation}
\begin{equation}
\Psi_2^{(i)}({\bf r})=\frac{\partial \Phi}{\partial n_2}+\frac{1}{4 \pi R_i}
\frac{\partial \Phi}{\partial n_1} + \frac{1}{4 \pi R_i^2}
\frac{\partial \Phi}{\partial n_0},
\end{equation}
and 
\begin{equation}
\Psi_{v2}^{(i)}({\bf r})=\frac{\partial \Phi}{\partial n_{v2}}+
\frac{1}{4 \pi R_i} \frac{\partial \Phi}{\partial n_{v1}}.
\end{equation}
These functions depend on the particular version of FMT employed. Here we use
the White-Bear version \cite{roth02a}.

The slit geometry that we consider in the present study allows us to further
simplify the gradient in Eq.~(\ref{grad}). The effective one-dimensional
weight functions and their derivatives with respect to $z$ are easily calculated.
For the volume weight function we find
\begin{equation}
\omega_3^i(z) = \pi (R_i^2 -z^2)~ \Theta(R_i-|z|),
\end{equation}
where $\Theta(z)$ denotes the Heaviside step-function. The derivative of the
volume weight function is
\begin{equation}
W_3^i(z) = \frac{\partial}{\partial z} \omega_3^i(z) = -2 \pi z~
\Theta(R_i-|z|).
\end{equation}
Note that the derivative of the Heaviside function can be neglected,
because the weight function $\omega_3^i(z)$ vanishes at the integration
boundaries. The effective surface weight function is
\begin{equation}
\omega_2^i(z) = 2 \pi R_i~ \Theta(R_i-|z|),
\end{equation}
which is constant in the range of integration, so that its derivative
\begin{equation}
W_2^i(z) = \frac{\partial}{\partial z} \omega_2^i(z) = 2 \pi R_i~
\left[\delta(z+R_i)- \delta(z-R_i)\right]
\end{equation}
has contributions only from the derivative of the Heaviside-$\Theta$ function.
Finally, the vector-like weight function is
\begin{equation}
\omega_{v2}^i(z) = -2 \pi z~ \Theta(R_i-|z|)~ {\bf e}_z,
\end{equation}
and leads to a derivative
\begin{eqnarray}
W_{v2}^i(z) &=& \frac{\partial}{\partial z} \omega_{v2}^i(z) = \left\{-2
\pi~\Theta(R_i-|z|) \right.  \\
&+& \left. 2 \pi R_i~
\left[\delta(z+R_i)+\delta(z-R_i)\right] \right\} ~ {\bf e}_z. \nonumber
\end{eqnarray}

Putting all the ingredients together we obtain the following expression
for the gradient of the variation of the excess free energy in planar geometry:
\begin{eqnarray}
\frac{\partial}{\partial z} \frac{\delta {\cal F}_{ex}}{\delta \rho_i(z)} 
& = & \int \left\{\Psi_3^{(i)}(z')~ W_3^i(z'-z) \right. \nonumber \\
& + & \left. \Psi_{v2}^{(i)}(z')~ W_{v2}^i(z'-z)\right\}~d z' \nonumber \\
& + & 2 \pi R_i \left\{ \Psi_2^{(i)}(z+R_i)-\Psi_2^{(i)}(z-R_i) \right. 
\nonumber \\
& + & \left.\Psi_{v2}^{(i)}(z+R_i)+\Psi_{v2}^{(i)}(z-R_i) \right\}.
\end{eqnarray}

\begin{acknowledgments}
A.J.A. acknowledges financial support from the British Council, funded under
the ARC programme, and also from RCUK\/. M.R. acknowledges 
financial support from the priority program SPP~1164 of the Deutsche
Forschungsgemeinschaft. M.R. and R.R. acknowledge financial support from DAAD
funded under the PPP programme. 
\end{acknowledgments}


\end{document}